%Paper: 9109040
%From: seiberg@seiberg.rutgers.edu
%Date: Mon, 23 Sep 91 10:04:39 EDT

\input harvmac

\def\IR{\relax{\rm I\kern-.18em R}}

\def\prl{Phys. Rev. Lett. }

\Title{\vbox{\baselineskip12pt\hbox{RU-91-41}}}
{{\vbox {\centerline{Irrational Axions as a Solution of}
\smallskip
\centerline{The Strong CP Problem in an Eternal Universe}
}}}
\bigskip

\centerline{\it Tom Banks$^{*}$, Michael Dine$^{**}$ and Nathan Seiberg$^{*}$}
\medskip
\centerline{$^{*}$Department of Physics and Astronomy}
\centerline{Rutgers University, Piscataway, NJ 08855-0849, USA}
\medskip
\centerline{$^{**}$Santa Cruz Institute for Particle Physics}
\centerline{University of California, Santa Cruz, CA 95064 USA}
\bigskip

\noindent
We exhibit a novel solution of the strong CP problem, which does not
involve any massless particles.  The low energy effective Lagrangian of
our model involves a discrete
spacetime independent axion field which can be thought of as a parameter
labeling a dense set of $\theta$ vacua.  In the full theory this
parameter is seen to be dynamical, and the model seeks the state of
lowest energy, which has $\theta_{eff} = 0$.  The processes which mediate
transitions between $\theta$ vacua involve heavy degrees of freedom and
are very slow.  Consequently, we do not know whether our model can solve
the strong CP problem in a universe which has been cool for only a
finite time.  We present several speculations about the
cosmological evolution of our model.

\Date{Sep. 1991}
%\draftmode

\newsec{Introduction and a 1+1 Dimensional Example}

There have been a number of mechanisms proposed for resolving the strong
CP problem of Quantum Chromodynamics (QCD).  Most of them are either
ruled out by experiment or only marginally consistent with the
combination of experimental data and conventional cosmology\foot{The
single exception that we are aware of is the class of models
\ref\dissy{A. Nelson, Phys.Lett. {\bf 136B} (1984) 387; S. Barr,
Phys. Rev. Lett. {\bf 53} (1984) 329, Phys. Rev. {\bf D30} (1984) 1805;
R. Mohapatra and K. Babu, Phys. Rev. {\bf D41} (1990) 1286}
in which discrete symmetries guarantee that the argument of the
determinant of the quark mass matrix is zero at tree level.}.  In the
present note we would like to present a new solution of the strong CP
problem.  We will exhibit a flat space-time quantum field theory in
which the QCD $\theta$ parameter is ``screened.''  That is to say,
although $\theta$ appears in the Lagrangian of the model, the observable
correlation functions in the true ground state do not depend on it.  In
particular, they are all CP invariant.

We do not yet know whether our model provides a realistic solution of
the strong CP problem in the world, as opposed to a model of a flat
eternal universe.  It has a large number of nearly degenerate metastable
vacuum states, and it is not clear that the system will ever find its
ground state in a universe which has been large only for a finite time.
The details of the discussion may also depend on the mechanism that
ensures that the cosmological constant vanishes.  Since our
understanding of these cosmological issues is not complete, we can only
offer a few speculations about them.  These will be presented in section III.

The idea for our solution of the strong CP problem is most easily
demonstrated by first examining another theory with a $\theta$
parameter: $1+1$ dimensional QED.  The Abelian gauge group of
the theory can be either the compact group $U(1)$ or its non-compact
covering group \IR.  The choice between \IR\ and $U(1)$ has significant
consequences.  The charges of the matter fields in the \IR\ theory are
not restricted but they have to be integers in the $U(1)$ theory.
Another difference between these two theories is that unlike the $U(1)$
gauge theory, the \IR\ gauge theory on a compact two-dimensional
parameter space has no $\theta$ dependence.  On a non-compact parameter
space both theories exhibit $\theta$ dependence which is interpreted as
a background electric field originating from classical charges at
infinity.  This fact remains true even in the presence of dynamical
matter fields whose charges are quantized.

Now consider the case of dynamical matter fields whose charges are not
quantized.  In particular, consider the Abelian Higgs model with Higgs
field of charge one and a massive fermion of irrational charge $q$.
It is particularly interesting to examine the limit in which the fermion
mass is much larger than the scale defined by the Higgs model.
Since the charges are not quantized, the gauge group must be \IR\ and
there is no $\theta$ dependence on a compact parameter space.  Perhaps
more surprising is the fact that there is also no $\theta$ dependence on
a non-compact parameter space.  The point is that any background
electric field $\theta$ can be screened by popping charges out of the
vacuum.  With the two charges, one and $q$, every value of
$\theta=2\pi(n+mq)$ with $n$ and $m$ integers can be screened.  The
screening process may cost a lot of energy (and be very slow) because
the irrationally charged fermions are so heavy, but this will always be
compensated by gaining the constant field energy over the infinite
volume of space. Since the ground state energy is minimized when the
external field is zero and since any value of $\theta\over 2\pi$ can be
arbitrarily closely approximated\foot{The fundamental result that is
needed to prove this is that the error in a rational approximant to any
number can be made to vanish like one over the square of the denominator
in the approximant
\ref\niven{I. Niven, {\it Numbers: Rational and
Irrational}, Random House, New York, p.94}.}
by a number of the form $(n + m q)$, the ground
state has vanishing background field regardless of the value of $\theta$
in the Lagrangian. This is the essence of our solution of the strong CP
problem.  The bosonized version of this model can be immediately
generalized to $3+1$ dimensions.  This will be done in the next section.

\newsec{Irrational Discrete Axions}

Consider a model consisting of an axion, $a$, coupled to two nonabelian gauge
fields in the standard manner.  The Euclidean Lagrangian has the form
\eqn\lag{{\cal L} = {f^2\over 2} (\partial a)^2 + {1\over 4 g_1^2} F_1^2 +
{1\over 4 g_2^2} F_2^2 + i (a +\theta_1) Q_1 + i (q a + \theta_2 ) Q_2}
where the $Q_i$ are the topological charge densities of the two gauge
fields, $f$ is the axion decay constant and $q$ is a dimensionless
parameter.  The second gauge group is supposed to represent QCD, while
the first is another confining theory with a much larger confinement
scale, $\Lambda_1 \gg \Lambda_2=\Lambda_{QCD}$.  We assume that the
fermions coupled to these two gauge theories have no global symmetries
that could be used to rotate away the topological terms in the
Lagrangian.  Clearly, without loss of generality we can use the shift
symmetry of the axion to set $\theta_1=0$.  The conventional wisdom is
that this leaves us
with no freedom to change the value of $\theta_2$ and the model
\lag\ has strong CP violation.  However, a closer examination shows that
after $\theta_1$ is set to zero we still have the freedom to shift $a$
by $2\pi n$ for integer $n$.  If the parameter $q$ is irrational, we can
use this freedom to set $\theta_2$ arbitrarily close to any desired
value.  Therefore, the theory based on the Lagrangian \lag\ is
independent of both $\theta_1$ and $\theta_2$ and is equivalent to
\eqn\lago{{\cal L} = {f^2\over 2} (\partial a)^2 + {1\over 4 g_1^2} F_1^2 +
{1\over 4 g_2^2} F_2^2 + i a (Q_1 + q Q_2) ~~.}
Unlike the conventional wisdom, one axion field $a$ can remove more than
one $\theta$ parameter.

What we have just argued is that the theory based on \lag\ is equivalent
to that based on \lago.  The latter is obviously CP invariant, if the
axion field $a$ is defined to change sign under a CP transformation.  What we
would like to show now is that CP is not spontaneously broken and
therefore all correlation functions are CP invariant.  To show that we
first integrate out all the non-zero modes of $a$.  The result of this
Gaussian integral is
\eqn\lage{{\cal L}_{eff} = {1\over 4 g_1^2} F_1^2(x) +
{1\over 4 g_2^2} F_2^2(x) + {q \over f^2}\int dy Q_1(x) G(x,y) Q_2(y) +
i a_0 (Q_1(x) + q Q_2(x)) }
where $G(x,y)$ is the propagator of $a$ with the zero mode, $a_0$,
removed.   Now, integrating over the gauge fields we find the effective
potential $V_{eff}(a_0)$.  Following the argument in
\ref\witvaf{C. Vafa and E. Witten, \prl {\bf 53} (1984) 535.},
we see that the point $a_0=0$ is a global minimum of $V_{eff}$. The
integrand of the gauge field functional integral is positive for $a_0 =
0$ and carries a phase for any other value of $a_0$.  The partition
function is therefore maximized, and the energy minimized, for the CP
conserving vacuum $a_0 = 0$.

Let us examine the theory in more detail in order to understand the
physical mechanism for resolving the strong CP problem.  We set
$\theta_1=0$ and integrate out the heavy gauge degrees of freedom.  This
fixes the VEV of the axion to $2\pi n$ for some integer $n$ and the
axion acquires a mass of order $\Lambda_1^2 \over f$.  Typically, one
ignores the integer $n$ and sets it to zero.  Then the low energy theory
includes the light gauge fields (QCD), no axion and a $\theta$ parameter
equal to $\theta_2$.  This is essentially the argument that one axion
can remove only one $\theta$ parameter.  However, the general argument
in the previous paragraphs shows that when $q$ is irrational the theory
cannot depend on $\theta_2$.  The reason for that is that the integer
$n$ cannot be ignored.  After integrating out the massive gauge fields,
the theory has an infinite number of degenerate ground states labeled by
$n$.  The dynamics of the light gauge fields breaks this degeneracy.
When $f\gg \Lambda_i$ the term with the propagator in \lage\ is small
and can be neglected.  In this approximation the effective potential
$V_{eff}(a_0)$ has the form
\eqn\effpot{V_{eff}(a) = \Lambda_1^4 E_1 (a) +\Lambda_2^4 E_2
(qa + \theta_2 ) }
where we have dropped the subscript of $a$.  The functions $E_i$ are
both periodic with period $2\pi$ and have their minimum when the
argument vanishes \witvaf.  Assuming that $E_i$ are continuous at $0~{\rm
mod}~2\pi$ we can minimize the total effective potential by making both
of the arguments of the $E_i$ as close as possible to multiples of
$2\pi$.  Thus
\eqn\mina{a \approx 2\pi n  \qquad ; \qquad qa+ \theta_2 \approx 2\pi m}
These two equations are compatible iff
\eqn\compat{\theta \approx 2\pi (m- qn )~~.}
These are of course just the equations that we discussed in the $1+1$
dimensional Higgs model.  If $q$ is irrational, then we can satisfy this
condition with arbitrary precision by appropriate choice of $n,m$.

Our mechanism has an obvious generalization to the case of several gauge
groups all coupled to the axion $a$ through some coefficients $q_i$.  If
all these coefficients $q_i$ are relatively irrational, there is an
infinite set of vacua where $(q_i \langle a\rangle + \theta_i) {\rm
mod} 2\pi < \epsilon_i $ for any $\epsilon_i$.  It might however be
important for cosmological reasons to note that the fraction of vacua
satisfying these inequalities goes like $\prod_i \epsilon_i$ when all
the $\epsilon_i$ are small.  Thus if the dynamics of the universe
randomly chooses between all possible metastable states of the system,
the probability of not seeing any low energy CP violation goes rapidly
to zero as the number of gauge groups is increased.

The failure of the decoupling theorem for this model stems from two
separate sources.  First of all, the high energy theory has an infinite
set of degenerate vacua, and the degeneracy is broken by QCD.  Equally
importantly, we are discussing the ground state of the model, which
means that we are willing to wait an arbitrarily long time for the
system to settle down.  The processes by which the system moves from one
of the almost degenerate metastable states with $\theta_{QCD} \neq 0$,
to the true vacuum will be very slow.  For the purposes of discussing
local physics over finite time intervals the decoupling theorem is
valid.

An equivalent way of thinking about the model is the following.  We can
include in our low energy effective Lagrangian a {\it discrete field}
$n$ which is independent of the coordinates, $x$, and takes values in
the integers.\foot{It should be stressed that at the level of the
low energy theory, this field has no dynamics.}
This field represents the value of $\langle a
\rangle/2\pi$ and thus labels the almost degenerate ground states.  It
is crucial that by varying the value of $n$ every value of $\theta_{QCD}
= (q \langle a \rangle +\theta_2)~{\rm mod}~2\pi$ can be approximated
arbitrarily well.  Therefore, the sum in the functional integral of the
low energy theory over $n$ can be replaced by an integral over a single
continuous variable $\theta_{QCD}$ which is independent of $x$.
Therefore, our theory looks like ordinary QCD with one more integration
variable $\theta_{QCD} $. The standard problem with such a theory is
that ordinarily different values of $\theta_{QCD} $ correspond to
different superselection sectors.  There are no physical processes or
local operators which communicate between these different sectors and
therefore $\theta_{QCD} $ should not be integrated over.  The novelty in
our theory is that $\theta_{QCD} $ does not label different
superselection sectors.  The high energy theory makes the barrier
between these sectors finite and allows transitions between them.

As with the standard axion, $\theta_{QCD} $ is a field which is
integrated over and can relax to zero.  In our case, though, only the
zero momentum mode of the field $n$ and therefore also only the zero
momentum mode of $\theta_{QCD} $ exists.  Hence, we do not have a
massless or light axion.

One might ask whether our model suffers from the $U(1)$ problem when
massless quarks are coupled to it.  Then it has an axial $U(1)$ symmetry
under which the coordinate $\theta_{QCD}$ is shifted by a
constant.  However, since $\theta_{QCD}$ is independent of $x$, it does
not have a conjugate momentum and no charge generates this symmetry.
{}From the high energy point of view this follows from the fact that
nearby values of $\theta_{QCD} $ are associated with far separated
points in field space.  Therefore, when the symmetry is broken there is no
Goldstone boson and there is no $U(1)$ problem.  A simple toy model
which exhibits such a behavior is based on the Lagrangian
\eqn\toy{{\cal L}_{toy}= {f_\eta^2 \over 2} \partial_\mu \eta(x)
\partial^\mu \eta(x) + V(\eta(x) + \theta_{QCD}) }
where $ \theta_{QCD}$ is an integration variable in the functional
integral.  The field $\eta(x)$ plays the role of the would be light
boson of the $U(1)$ problem.  The theory \toy\ is invariant under the
broken $U(1)$ symmetry $\eta(x) \rightarrow \eta(x) + \alpha$;
$\theta_{QCD} \rightarrow \theta_{QCD} - \alpha$ and the potential $V$
can be arbitrary.  We can use the $U(1)$ symmetry to set $\theta_{QCD}
=0$ in the Lagrangian and then the integral over $\theta_{QCD} $
factorizes.  Clearly, the resulting theory does not have a massless
$\eta$ particle.

Technically, the violation of current algebra \lq\lq
theorems'' relating the $U(1)$ Goldstone boson to $\theta$ dependence in
QCD comes about in our model because the two point function of
topological charge density is discontinuous at zero momenta.  The
discontinuity comes from intermediate metastable vacuum states, which
are exactly stable in the low energy theory.  We repeat that
in the low energy theory we sum over superselection sectors.  This is
\lq\lq forbidden'' by the \lq\lq axioms'' of field theory, but in our
model the high energy sector provides a dynamical rationale for summing
over these sectors.  The full theory satisfies all relevant axioms.

\newsec{Cosmological Speculations}

The model that we have discussed so far would solve the strong CP
problem in an eternal flat world in the absence of gravitation.  In the
real world, an expanding universe which has undoubtedly been at a
temperature below the QCD scale for only a finite time, one must ask
whether the system we have described will ever find its true ground
state.  The answer to this question may involve very complicated,
perhaps chaotic or spin glass-like dynamics, and/or be connected to other deep
cosmological puzzles.  At the present time we have no clear picture of
how the model behaves. We will therefore simply suggest some possible
scenarios and leave a more serious discussion for future work.

The first scenario is the simplest, and adheres most closely to the
standard discussion of axion cosmology.  It should be applicable for at
least some range of parameters in our model.  In this standard scenario
we assume that the
axion field is sufficiently weakly coupled that after inflation it
simply falls into one of its classical vacua over regions much larger
than the entire universe visible to us today.  It is easy to argue that
when $\theta_{eff}=(qa+\theta_2)~{\rm mod}~2\pi$ is very small, the
fraction of vacua with $\theta < \theta_{eff}$ is linear in
$\theta_{eff}$, so there is only one chance in $10^{-9}$ that any given
region has $\theta$ small enough to be compatible with the current
experimental bound on the neutron electric dipole moment.  All is not
lost however if we make the further assumption that the cosmological
constant vanishes at the true minimum of the axion potential.  We
emphasize that we have no idea why this is so, but that this is the
standard assumption about the cosmological constant.

Given these assumptions, the cosmology of our model is fairly standard
until temperatures of order $T_c \sim \theta^{\ha} \Lambda_{QCD}$.  In
particular, since the axion potential is of order $\Lambda_H^4$, the
energy density is not dominated by nonrelativistic axions which
overclose the universe.\foot{At least, it is possible to choose a wide
range of parameters for which the axion lifetime is short enough that
 this problem does not arise.}
  Instead, cosmology follows the standard
Robertson Walker scenario until the universe reaches temperatures of
order $T_c$.  At this point, the universe becomes cosmological constant
dominated, with a cosmological constant $\sim \theta^2 \Lambda_{QCD}^4$.
Weinberg
\ref\weinberg{S. Weinberg, \prl {\bf 59} (1987) 2607.}
has shown that a positive cosmological constant greater than about
$10^3$ times the present observational limit will prevent the formation
of galaxies.  Thus, the only regions in which galaxy formation can take
place are those in which
\eqn\thetabnd{\theta^2 \Lambda_{QCD}^4 \leq 10^{- 9} eV^4}
Since $\Lambda_{QCD} \sim 10^8 eV$, regions containing galaxies have
$\theta \leq 10^{- 20}$.
In other words, conventional inflationary
axion dynamics, coupled with the standard assumption that the
cosmological constant vanishes at the absolute minimum of the potential,
implies that the only regions in our model universe which contain
galaxies are those with $\theta$ much smaller than the bound from the
neutron electric dipole moment.  This seems to us to be a reasonably
attractive resolution of the strong CP problem.  Perhaps its greatest
defect is that only about $1 /30$ of the metastable domains containing
galaxies will have a cosmological
constant consistent with the present limit (despite the fact that we
have fine tuned the true ground state energy to zero).

We are not at all sure that our model behaves as we have described in
the previous paragraph.
Our alternative scenarios are harder to analyze and none seem to work
very well.  They involve the assumption that the presently observable
piece of the universe consists of multiple domains in which the
effective value of $\theta$ is different.  One must then analyze the
dynamics of these domains as the temperature falls below the QCD scale.
Those with very small values of $\theta$ are energetically favored, and
begin to expand relative to the others.\foot{Here we assume that the
bubbles are larger than their critical size when the QCD temperature is
reached. The critical size is quite large because the surface tension in
the domain walls is determined by the heavy scale $\Lambda_H$.} On the
other hand, those with higher values of $\theta$ become dominated by
their cosmological constants and expand exponentially.  There are
possible contributions to the energy density from domain walls, and
nonrelativistic axion gases. The situation is
made more complicated by the bizarre nature of the potential.  States
that are close in energy are far away in field space.  Even a single
classical variable with such a potential has chaotic behavior
and we are dealing with a field theory
full of such degrees of freedom.  At the present time we believe that
the expansion of large $\theta$ domains due to their cosmological
constants is the dominant effect.  It is hard to see how a universe
built in such a manner could resemble our own.  Nonetheless, we feel
that these complicated scenarios should be understood more fully.  The
wild speculation that the axion domains in such a chaotic system might
have something to do with the foamlike large scale structure that has
been recently observed
\ref\foam{H. Rood, Ann. Rev. Astron. Astrophys. {\bf 26} (1988) 245, and
references cited therein.}
is immensely attractive.  Indeed, because of the scarcity of states with
small $\theta$ one might imagine that the most probable multiple domain
configurations with galaxies would have some domains just above and some
just below the Weinberg bound.  It is amusing
to speculate that the famous
voids in Bootes and other parts of the sky are regions in which the
effective cosmological constant was too large to allow for galaxy
formation.  For lack of talent and insight, we will have to leave such
cosmic fantasies for a future publication.

Another speculative application of the irrational axion idea
is to an
anthropic solution of the cosmological constant problem\foot{Some
time ago, L. Abbott
\ref\abbott{L. Abbott,
Phys. Lett. {\bf 150B} (1985) 427.} suggested a scheme
for cancelling the cosmological constant also involving axions.
Unlike the present proposal, it was necessary to introduce
an extremely small energy scale, and the low energy theory
contained a light particle.}.
Imagine that
in some version of supergravity it is natural for the cosmological
constant to be at the SUSY breaking scale M.  Now consider the SUSY
version
of the model of this paper with both gauge groups also at the SUSY
breaking scale.  The total effective potential is $M^4(K + F(a))$.
$F(a)$
has a set of minima with energies that fill an interval of order 1
densely.  Thus there are many vacua in which K is cancelled to an
accuracy sufficient to allow galaxy formation,
and in a fraction $ 10^{-3}$
of those, the cosmological constant is as small as that observed in our
universe.
A typical state with small cosmological constant
will not be sufficiently metastable to serve as a model
for our universe, however some fraction of {\it these} vacua will be.

\newsec{Conclusions}

In this paper we provide an example of a flat space field theory which
solves the strong CP problem without massless particles of any kind.
The model violates the usual decoupling theorems in an interesting way.
The low energy theory has a discrete global variable, $n$, labeling a
set of quasi degenerate vacua.  The dynamics that allows these vacua to
transform into one another and settle down into the true ground state
cannot be understood without appealing to the high energy theory.  The
decoupling theorems are still valid in the weak sense that local
dynamics in each metastable ground state is described completely in
terms of the low energy Lagrangian.  The discrete variable does allow us
to evade the usual argument connecting the strong CP problem to the
$U(1)$ problem.

On the negative side, the basic Lagrangian of our model
is nonrenormalizable, and it is easy to show that it cannot be the
effective theory of any renormalizable field theory.  In addition we
have not been able to find a string compactification which leads to an
axion with such irrational couplings.  Thus, the fundamental basis for
our model remains obscure.  A possible origin for the irrational
couplings central to our model may be found in the novel nonperturbative
behavior of string theory that has been pointed out by
Shenker
\ref\shenker{S. Shenker, in the Proc. of the
Cargese meeting, Random Surfaces, Quantum Gravity and Strings, (1990).}.
He argued that intrinsically stringy
nonperturbative effects will behave as $e^{- {1\over g}}$ instead of the
$e^{-{1\over g^2}}$ characteristic of field theory, and has speculated
that this behavior could be understood in terms of \lq\lq instantons of
continuous topological charge''
\ref\shenkspec{S. Shenker, {\it Private Communication.}}.
If such instantons indeed exist in string theory, they
might provide the irrational axion couplings that we require.

The general features of our mechanism may be applicable to other fine
tuning problems in particle physics.  Its fundamental characteristic is
that it allows us to turn couplings into dynamical variables without
invoking massless particles\foot{Or wormholes!}.  One need only have a
high energy field with a discrete set of degenerate ground states whose integer
label is irrationally related to a parameter in the low energy
Lagrangian (so that the parameter
can be given a dense set of values by appropriate
choice of the integer).  Low energy dynamics resolves the degeneracy and
high energy processes mediate the transitions between states with
different values of the effective coupling.  We have already described a
crude version of how such a mechanism might help us to understand the
cosmological constant problem.  It is to be hoped that one can do better
than this.  In string theory, our mechanism might help to resolve the
problem of determining the string coupling.  Conventionally it is said
that any
potential which allows the fine structure constant to be weak, and does
not force it to vary significantly over geological time scales, implies
the existence of a scalar particle of very small mass
\ref\ds{M. Dine and N. Seiberg, Phys. Lett. {\bf 162B} (1985) 299.}.
Since string theory also determines the couplings of this particle to be
about gravitational strength, it
is ruled out by astrophysical considerations.  We now envisage the
possibility of generating an \lq\lq irrational'' potential for the
string coupling, which could determine it (or allow it to be set as an
initial condition for our part of the universe) without requiring any
massless particles.  This exciting idea is also left for future work.

We do not know whether we have found a solution to the strong CP problem
in the real world.  Conventional assumptions about axion dynamics in an
inflationary cosmology, and about the value of the cosmological
constant, lead to a correlation between the existence of galaxies in the
observable part of the universe and the fact that $\theta$ is so small.
Alternative assumptions about the spatial configuration of the axion
field in our universe might lead to an explanation of foam like large
scale structure, and great voids.  At present, these two pictures do not
seem compatible with each other, and the second probably leads to a
highly inhomogeneous universe.  However, our present understanding of
the cosmology of this model is such that we can still hope for the best
of all possible worlds.

\bigskip
\centerline{\bf Acknowledgements}

It is a pleasure to thank S. Coleman,
M. Douglas, D. Friedan, E. Martinec, G. Moore,
R. Leigh, S. Shenker, L. Susskind and A.B. Zamolodchikov
for several useful discussions.  This
work was supported in part by DOE grants DE-FG05-90ER40559
and DE-AM03-76SF00010.

\listrefs
\end